\documentclass[twocolumn,letterpaper,aps,prl,superscriptaddress,showpacs,nofootinbib,floatfix]{revtex4}

\usepackage{graphicx}
\usepackage{slashbox}
\usepackage{brackets}
\usepackage{psfrag}
\usepackage{tabularx}
\begin{document}

\ProvideTextCommandDefault{\textonehalf}{${}^1\!/\!{}_2\ $}

\title{First Measurements of Spin-Dependent Double-Differential Cross Sections and the GDH Integrand from $\vec{{^3}He}(\vec{\gamma},n)pp$ at Incident Photon Energies of 12.8 and 14.7 MeV}

\author{G.~Laskaris}\email[Electronic address: ]{georgios.laskaris@duke.edu}
\affiliation{Triangle Universities Nuclear Laboratory, Durham, North Carolina 27708, USA}
\affiliation{Department of Physics, Duke University, Durham, North Carolina 27708, USA}
\author{Q.~Ye}
\altaffiliation[Currently at ]{UMD, College Park, MD 20742, USA}
\affiliation{Triangle Universities Nuclear Laboratory, Durham, North Carolina 27708, USA}
\affiliation{Department of Physics, Duke University, Durham, North Carolina 27708, USA}
\author{B.~Lalremruata}
\altaffiliation[Currently at ]{Mizoram University, Aizawl 796004, India}
\affiliation{Triangle Universities Nuclear Laboratory, Durham, North Carolina 27708, USA}
\affiliation{Department of Physics, Duke University, Durham, North Carolina 27708, USA}
\author{Q.~J.~Ye}
\affiliation{Triangle Universities Nuclear Laboratory, Durham, North Carolina 27708, USA}
\affiliation{Department of Physics, Duke University, Durham, North Carolina 27708, USA}
\author{M.~W.~Ahmed}
\affiliation{Triangle Universities Nuclear Laboratory, Durham, North Carolina 27708, USA}
\affiliation{Department of Physics, Duke University, Durham, North Carolina 27708, USA}
\affiliation{Department of Mathematics and Physics, North Carolina Central University, Durham, North Carolina 27707, USA}
\author{T.~Averett}
\affiliation{College of William and Mary, Williamsburg, Virginia 23187, USA}
\author{A.~Deltuva}
\affiliation{Centro de F\'{i}sica Nuclear da Universidade de Lisboa, P-1649-003 Lisboa, Portugal}
\author{D.~Dutta}
\affiliation{Mississippi State University, Mississippi State, Mississippi 39762, USA}
\author{A.~C.~Fonseca}
\affiliation{Centro de F\'{i}sica Nuclear da Universidade de Lisboa, P-1649-003 Lisboa, Portugal}
\author{H.~Gao}
\affiliation{Triangle Universities Nuclear Laboratory, Durham, North Carolina 27708, USA}
\affiliation{Department of Physics, Duke University, Durham, North Carolina 27708, USA}
\author{J.~Golak}
\affiliation{M. Smoluchowski Institute of Physics, Jagiellonian University, PL-30059 Krak\'{o}w, Poland}
\author{M.~Huang}
\affiliation{Triangle Universities Nuclear Laboratory, Durham, North Carolina 27708, USA}
\affiliation{Department of Physics, Duke University, Durham, North Carolina 27708, USA}
\author{ H.~J.~Karwowski}
\affiliation{Triangle Universities Nuclear Laboratory, Durham, North Carolina 27708, USA}
\affiliation{Department of Physics and Astronomy, University of North Carolina at Chapel Hill, Chapel Hill, North Carolina 27599, USA}
\author{J.~M.~Mueller}
\affiliation{Triangle Universities Nuclear Laboratory, Durham, North Carolina 27708, USA}
\affiliation{Department of Physics, Duke University, Durham, North Carolina 27708, USA}
\author{L.~S.~Myers}
\affiliation{Triangle Universities Nuclear Laboratory, Durham, North Carolina 27708, USA}
\affiliation{Department of Physics, Duke University, Durham, North Carolina 27708, USA}
\author{C.~Peng}
\affiliation{Triangle Universities Nuclear Laboratory, Durham, North Carolina 27708, USA}
\affiliation{Department of Physics, Duke University, Durham, North Carolina 27708, USA}
\author{B.~A.~Perdue}
\altaffiliation[Currently at ]{LANL, Los Alamos, NM 87544, USA}
\affiliation{Triangle Universities Nuclear Laboratory, Durham, North Carolina 27708, USA}
\affiliation{Department of Physics, Duke University, Durham, North Carolina 27708, USA}
\author{X.~Qian}
\altaffiliation[Currently at ]{Caltech, CA 91125, USA}
\affiliation{Triangle Universities Nuclear Laboratory, Durham, North Carolina 27708, USA}
\affiliation{Department of Physics, Duke University, Durham, North Carolina 27708, USA}
\author{P.~U.~Sauer}
\affiliation{Institut f\"ur Theoretische Physik, Leibniz Universit\"at Hannover, D-30167 Hannover, Germany}
\author{R.~Skibi\'nski}
\affiliation{M. Smoluchowski Institute of Physics, Jagiellonian University, PL-30059 Krak\'{o}w, Poland}
\author{S.~Stave}
\altaffiliation[Currently at ]{PNNL, Richland, WA 99352, USA}
\affiliation{Triangle Universities Nuclear Laboratory, Durham, North Carolina 27708, USA}
\affiliation{Department of Physics, Duke University, Durham, North Carolina 27708, USA}
\author{J.~R.~Tompkins}
\altaffiliation[Currently at ]{MSU, East Lansing, MI 48824, USA}
\affiliation{Triangle Universities Nuclear Laboratory, Durham, North Carolina 27708, USA}
\affiliation{Department of Physics and Astronomy, University of North Carolina at Chapel Hill, Chapel Hill, North Carolina 27599, USA}
\author{H.~R.~Weller}
\affiliation{Triangle Universities Nuclear Laboratory, Durham, North Carolina 27708, USA}
\affiliation{Department of Physics, Duke University, Durham, North Carolina 27708, USA}
\author{H.~Wita{\l}a}
\affiliation{M. Smoluchowski Institute of Physics, Jagiellonian University, PL-30059 Krak\'{o}w, Poland}
\author{Y.~K.~Wu}
\affiliation{Triangle Universities Nuclear Laboratory, Durham, North Carolina 27708, USA}
\affiliation{Department of Physics, Duke University, Durham, North Carolina 27708, USA}
\author{Y.~Zhang}
\affiliation{Triangle Universities Nuclear Laboratory, Durham, North Carolina 27708, USA}
\affiliation{Department of Physics, Duke University, Durham, North Carolina 27708, USA}
\author{W.~Zheng}
\affiliation{Triangle Universities Nuclear Laboratory, Durham, North Carolina 27708, USA}
\affiliation{Department of Physics, Duke University, Durham, North Carolina 27708, USA}

\date{\today}

\begin{abstract}
The first measurement of the three-body photodisintegration of longitudinally-polarized $^3$He with a circularly-polarized $\gamma$-ray beam was carried out at the High Intensity $\gamma$-ray Source (HI$\gamma$S) facility located at Triangle Universities Nuclear Laboratory (TUNL). The spin-dependent double-differential cross sections and the contributions from the three-body photodisintegration to the $^3$He GDH integrand are presented and compared with state-of-the-art three-body calculations at the incident photon energies of 12.8 and 14.7 MeV. The data reveal the importance of including the Coulomb interaction between protons in three-body calculations.
\end{abstract}


\maketitle
The study of three-nucleon systems has long been of fundamental importance to nuclear physics~\cite{glockle1,Carlson}. Calculations using mainly the machinery of Faddeev~\cite{fad} and Alt-Grassberger-Sandhas equations (AGS)~\cite{Alt} have been carried out for three-body systems using a variety of nucleon-nucleon (NN) potentials~\cite{Stoks,machle}, and three-nucleon forces (3NFs) like Urbana IX (UIX)~\cite{Carl} or CD Bonn + $\Delta$~\cite{Deltuva0}, with the latter yielding an effective 3NF through the $\Delta$-isobar excitation.

Calculations for the three-body photodisintegration of $^3$He with double polarizations have been carried out. The calculations by Deltuva {\it et al.} are based on AGS equations and employ the CD Bonn + $\Delta$ potential~\cite{Deltuva0} with the corresponding single-baryon and meson-exchange electromagnetic currents (MEC) plus relativistic single-nucleon charge corrections. The results are obtained using the computational technology of Ref. ~\cite{Deltuva2}. The proton-proton Coulomb force is included using the method of screening and renormalization~\cite{Deltuva3}. Skibi\'nski {\it et al.} solve the Faddeev equations by using the AV18 potential and the UIX 3NF~\cite{Carl} accounting for single nucleon currents and the two most important MEC, the seagull and pion-in-flight terms. Their results are obtained using the methods described in Ref.~\cite{Skibinski}.

Recent advances in high intensity polarized beams and polarized $^3$He targets allow for tests of new spin-dependent observables predicted by theory. A polarized $^3$He target is often used as an effective polarized neutron target to extract the electromagnetic form factors~\cite{Gao,Xu,Riordan} and the spin structure functions~\cite{Anthony} of the neutron since the nuclear spin of $^3$He is carried mostly by the unpaired neutron. To acquire the information about the neutron using a polarized $^3$He target, nuclear corrections relying on the state-of-the-art three-body calculations need to be validated by experiments. While data from electrodisintegration of polarized $^3$He~\cite{Xiong} were used to test three-body calculations~\cite{golak}, data from polarized photodisintegration of $^3$He below the pion production threshold did not exist prior to this work.

The spin-dependent total cross sections from the three-body photodisintegration of $^3$He below pion production threshold are of further importance for the investigation of the Gerasimov-Drell-Hearn (GDH) sum rule~\cite{Drell}. The GDH sum rule relates the energy-weighted difference of the spin-dependent total photoabsorption cross sections $\sigma^P$ (for target spin and beam helicity parallel) and $\sigma^A$ (for target spin and beam helicity anti-parallel) to the anomalous magnetic moment of the target (nuclei or nucleons) as follows:

\begin{equation}
I^{GDH} = \int_{\nu _{thr}}^{\infty}(\sigma^P- \sigma^A)
{\frac{d\nu}{\nu}} = \frac{4\pi^{2}e^{2}}{M^{2}}\kappa^{2} I,
\label{Igdhr}
\end{equation}
where $\nu$ is the photon energy, $\nu_ {thr}$ is the pion production (two-body break-up) threshold on the nucleon (nucleus), $\kappa$ is the anomalous magnetic moment, $M$ is the mass and $I$ is the spin of the nucleon or the nucleus.

For the case of $^3$He, the energy range that interests us is from the two-body breakup threshold ($\sim$5.5 MeV), to the pion production threshold ($\sim$140 MeV)~\cite{gaoproc}. The aforementioned calculations~\cite{Deltuva2,Skibinski} demonstrate that the three-body breakup channel below 40 MeV dominates the integrand~\cite{gaoproc}. Therefore, a spin-dependent study of $\vec{^3He}(\vec{\gamma},n)pp$ not only provides a stringent test of the modern three-body calculations, but also serves as an important step towards an experimental test of the GDH sum rule on the $^3$He nucleus in the future when one combines measurements from $^3$He above the pion production threshold from other laboratories~\cite{Amarian}.

In this letter, we present the first measurement of spin-dependent double differential cross sections and the GDH integrand of $^3$He from three-body breakup using a longitudinally-polarized $^3$He target at incident photon energies of 12.8 and 14.7 MeV. The experiment was carried out at the HI$\gamma$S facility~\cite{higsreview} using a nearly monoenergetic, $\sim$100\% circularly-polarized $\gamma$-ray beam. The beam was pulsed at a rate of 5.5 MHz with intensities of 1~-~2$\times10^{8}\gamma/s$, having an energy spread of $\Delta \nu/\nu$=3.0\% at $\nu$=12.8 MeV and 5.0\% at $\nu$=14.7 MeV. The photon flux was monitored utilizing a 4.7 cm long D$_2$O cell and two BC-501A-based liquid scintillator neutron detectors placed transverse to the beam direction and detecting the neutrons from the deuteron photodisintegration process. The integrated photon flux was extracted based on the well-known deuteron cross sections~\cite{bernabei,blackston,birenbaum}.

The polarized $^3$He target cell used in the experiment was a one-piece Pyrex glassware with Sol-Gel coating~\cite{Brinker} which consisted of a spherical pumping chamber 8.1 cm in diameter and a cylindrical target chamber 39.6 cm long and 2.9 cm in diameter, connected by a transfer tube 8 mm in diameter and 9.6 cm long. The $^3$He filling density of the target was determined to be 6.5$\pm$0.1 amg. The pumping chamber of the cell, which was heated to $\sim$200 C$\,^{\circ}$, contained a mixture of Rb and K necessary for spin exchange optical pumping~\cite{Happer}. Circularly-polarized laser light at 794.8 nm polarized Rb atoms in the pumping chamber. The Rb atoms in turn transferred their polarization to $^3$He nuclei through spin-exchange collisions between Rb-K, Rb-$^3$He and K-$^3$He. To improve the optical pumping efficiency, a small quantity of N$_{2}$ (0.1 amg) was added into the cell as a buffer gas. A 20 G magnetic field, provided by a pair of Helmholtz coils $\sim$170 cm in diameter, defined the direction of the $^3$He nuclear polarization.  More details about this target can be found in~\cite{Kramer,Ye}. A N$_2$-only reference cell with the same dimensions as those of the $^3$He target and filled with the same amount of N$_2$ gas was employed for measuring backgrounds. In order to extract the spin dependent double differential cross sections and form the GDH integrand, $(\sigma^P-\sigma^A)/\nu$, the spin of the target was flipped every 15 min. The beam helicity was flipped only once towards the end of the experiment. The target polarization P$_t$ was measured using the nuclear magnetic resonance-adiabatic fast passage technique~\cite{Lorenzon}, which was calibrated daily using the electron paramagnetic resonance~\cite{epr4} technique to extract the absolute polarization. The polarization of the target throughout the experiment was between 38\% and 43\%.

Neutrons from the $\vec{{^3}He}(\vec{\gamma},n)pp$ process were detected using sixteen BC-501A-based detectors positioned 1 m away from the center of the $^3$He cell. The detectors were placed symmetrically on each side of the beam axis at laboratory angles of $30\,^{\circ}$, $45\,^{\circ}$, $75\,^{\circ}$, $90\,^{\circ}$, $105\,^{\circ}$, $135\,^{\circ}$, $150\,^{\circ}$ and $165\,^{\circ}$. The pulse height (PH), the time of flight (TOF), and the pulse shape discrimination (PSD)~\cite{mesytec} between photons and neutrons were recorded for each event. The outgoing neutron energy was determined using the measured TOF assuming the neutrons were emitted from the center of the $^3$He target cell.

The double-differential cross section for target spin parallel/anti-parallel to the beam direction is defined as
\begin{equation}
\frac{d^{3}\sigma^{P/A}}{d\Omega dE_{n}}=\frac{Y_{i,ext}^{P/A}}{\varepsilon^{syst}_{i} \Delta \Omega \Delta E N_{t} }
\label{Igdhr3}
\end{equation}
where $Y_{i,ext}^{P/A}=\frac{1}{2}(Y_{i}^{P}(1\pm \frac{1}{P_tP_b})+Y_{i}^{A}(1\mp \frac{1}{P_tP_b}))$ is the extracted normalized yield (neutron counts/integrated photon flux, N$_\gamma$) of $^3$He at the $i^{th}$ energy bin with $Y_{i}^{P/A}=Y_{i}^{P/A,^3He}-Y_{i}^{N_2}$ being the measured yield from the $^3$He cell after the subtraction of the N$_2$ reference cell background yield for both parallel and antiparallel states, $P_t$ and $P_b$ is the target and the beam polarization, respectively, $\varepsilon^{syst}_{i}$ is the system efficiency accounting for both the intrinsic efficiency of the neutron detector and the neutron multiple scattering effect calculated at the $i^{th}$ energy bin, $\Delta\Omega$ is the acceptance of the neutron detector, $\Delta E$ is the width of the neutron energy bin, and $N_{t}$ is the $^3$He target thickness determined to be (8.4$\pm$0.1)$\times$10$^{21}$ cm$^{-2}$. The system efficiencies $\varepsilon^{syst}_{i}$ were calculated as a function of E$_n$ using a GEANT4~\cite{geant} simulation of the experiment and the light-output response of the neutron detectors as simulated in Ref.~\cite{Trotter}.

The selection of the neutron events from $^3$He was based on cuts on the PH, TOF, and PSD values. A PSD cut was first applied to the $^3$He target data to remove photon events. Then, a PH cut was applied to determine the detector efficiency. The same cuts were applied to the data taken with the N$_2$ reference cell to subtract the background. The neutron detection efficiency varies rapidly as a function of neutron energy below 1.5 MeV~\cite{Alex,perdue}. Therefore, we report cross sections only for neutrons with energies above 1.5 MeV as defined by the TOF cut.

There were two types of systematic uncertainties in this experiment: bin-dependent and bin-independent uncertainties. The bin-dependent systematic uncertainties were in principle asymmetric as they arose from the PH cuts on the neutron spectra. These uncertainties affected the shape of the observed distributions. The bin-independent ones were symmetric and the major contributors were the detector efficiency (2.8\%)~\cite{Trotter1,setze}, the $^3$He target thickness (1.3\%), the detector acceptance (2\%) and the flux determination (5.7\%), in which the main contribution was from the D$_2$ photodisintegration cross section uncertainty (4.6\%)~\cite{bernabei,blackston,birenbaum}. The uncertainty of neutron energy, E$_n$, varied from 1~-~8\% depending on the detector angle and the outgoing neutron energy. The systematic uncertainty of the target and the beam polarizations are 5.5\% and 5\%, respectively.

\begin{figure}[h!]
  \centering
    \includegraphics[width=0.5\textwidth]{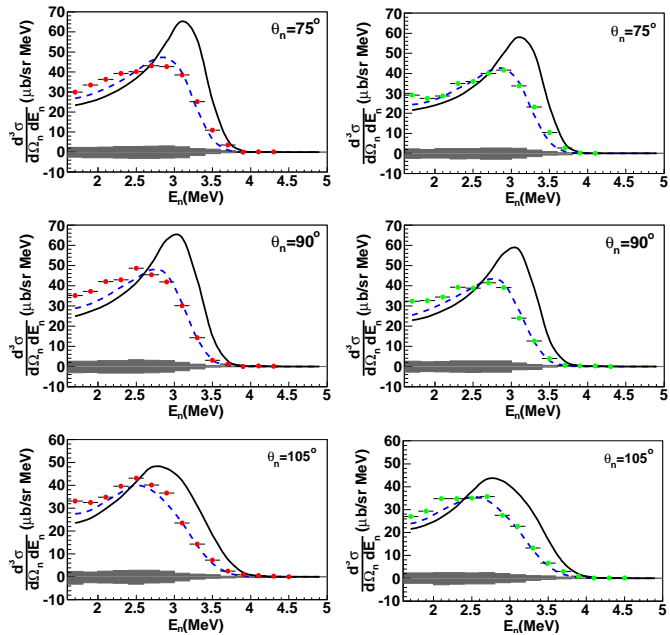}
    \caption{(Color online) Experimental spin-dependent double-differential cross sections for parallel (left panel) and antiparallel (right panel) states as a function of the neutron energy E$_n$ at
    $\nu$=12.8 MeV compared with the calculations of Deltuva {\it et al.} (dashed curve) and Skibi\'nski {\it et al.} (solid curve). The bin width is 0.2 MeV. The band shows the combined systematic uncertainties.}
   \label{fig:128}
\end{figure}

\begin{figure}[h!]
  \centering
    \includegraphics[width=0.5\textwidth]{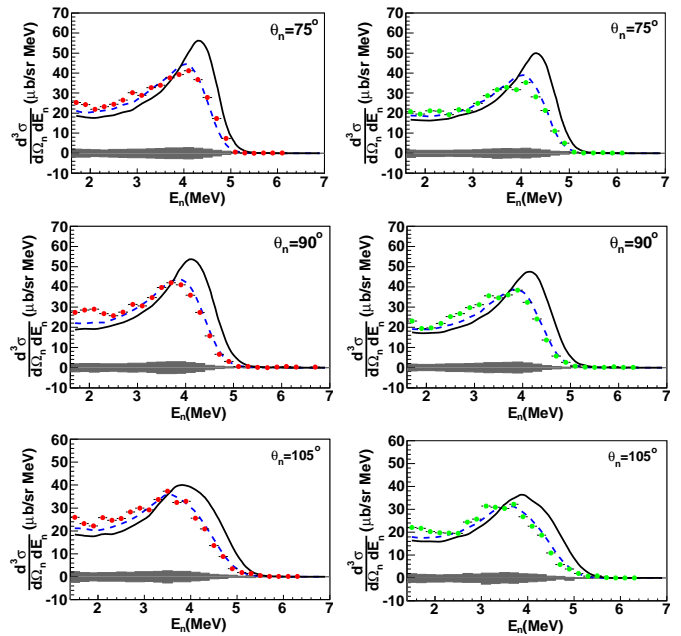}
    \caption{(Color online) As in Fig.~\ref{fig:128} but for $\nu$=14.7 MeV.}
    \label{fig:147}
\end{figure}

Figs.~\ref{fig:128} and~\ref{fig:147} show the spin-dependent double-differential cross sections at an incident photon energy of 12.8 and 14.7 MeV respectively, for parallel and antiparallel states as a function of the neutron energy at lab angles of $75\,^{\circ}$, $90\,^{\circ}$, and $105\,^{\circ}$. The dashed and solid curves are the GEANT4 simulation results using as cross section inputs the calculations provided by Deltuva {\it et al.} and Skibi\'nski {\it et al.} using the computational technology of Refs.~\cite{Deltuva2} and~\cite{Skibinski}, respectively. The band in each panel shows the overall systematic uncertainties combined in quadrature. The spin-dependent double differential cross sections for the rest of the scattering angles will be presented in a future publication.

The overall shape, magnitude, and location of the neutron peak in the experimental results are described better by the calculations of Deltuva {\it et al.} Further, studies show that the differences between Deltuva {\it et al.} and Skibi\'nski {\it et al.} are dominated by the proton-proton Coulomb force that is included only in the calculations by Deltuva {\it et al.} with all other ingredients playing a minor role in these differences. Therefore, one can conclude that the inclusion of the proton-proton Coulomb repulsion in the calculations is important for this process.

By integrating the double differential cross section distributions over the neutron energy, the partial differential cross sections for E$_n$$\textgreater$1.5 MeV can be extracted. The unmeasured part was added heuristically based on the average of the theoretical values below 1.5 MeV taken from Deltuva {\it et al.} and Skibi\'nski {\it et al.} The difference between the calculations is 1~-~8\% depending on the incident photon beam energy, the total spin-helicity state and the scattering angle, which introduces an additional systematic uncertainty to the differential cross sections of no more than 4\%. Legendre polynomials up to the 4$^{th}$ order were used to fit the differential cross sections. The fitting curves were integrated over the angle and the total cross sections were then extracted for both energies and the two spin-helicity states, in order to determine the values of the GDH integrand. The systematic uncertainties of the total cross sections were determined by varying the differential cross sections from the central values by plus or minus the overall systematic uncertainties and then performing the fit. The systematic uncertainty was taken as half of the difference between the two integrals of these two new fits. Details of this analysis and the results on the differential cross sections and asymmetries will be reported in a future publication.

Table~\ref{table:gdh} summarizes the spin-dependent total cross sections and the contributions from the three-body photodisintegration to the $^3$He GDH integrand for both photon energies and predictions from Deltuva {\it et al.} and Skibi\'nski {\it et al.} Better agreements between data and results from Deltuva {\it et al.} are again observed. 
The difference $\sigma^{P}-\sigma^{A}$ is sensitive not only to Coulomb repulsion, but also to relativistic single-nucleon charge corrections as already found in
Ref.~\cite{Deltuva2}.

\begin{table}[!ht]
\begin{center}
\caption{Total cross sections, $\sigma^{P}$ and $\sigma^{A}$, and the contributions from the three-body photodisintegration to the $^3$He GDH integrand, $(\sigma^{P}-\sigma^{A})/\nu$,  with statistical uncertainties followed by systematics, compared with theoretical predictions.}
\begin{tabular*}{0.48\textwidth}{@{\extracolsep{\fill} }c c c c }
\hline
\hline
\backslashbox{$\nu$ (MeV)}{ } & $\sigma^{P}$($\mu$b) & $\sigma^{A}$($\mu$b) & $(\sigma^{P}-\sigma^{A})/\nu$ (fm$^{3}$) \\ \hline
This work 12.8 & 861$\pm$5$\pm$81 & 765$\pm$5$\pm$71 & 0.147$\pm$0.010$\pm$0.018 \\ %\hline
Deltuva {\it et al.} & 872 & 777 & 0.146 \\ %\hline
Skibi\'nski {\it et al.} & 956  & 872 & 0.131 \\ %\hline
This work 14.7 & 999$\pm$5$\pm$89 & 869$\pm$5$\pm$78 & 0.174$\pm$0.011$\pm$0.020 \\ %\hline
Deltuva {\it et al.} & 1026 & 900 & 0.168 \\ %\hline
Skibi\'nski {\it et al.} & 1079 & 970 & 0.146 \\ \hline
\hline
\end{tabular*}
\label{table:gdh}
\end{center}
\end{table}

\begin{figure}[h!]
  \centering
    \includegraphics[width=0.5\textwidth]{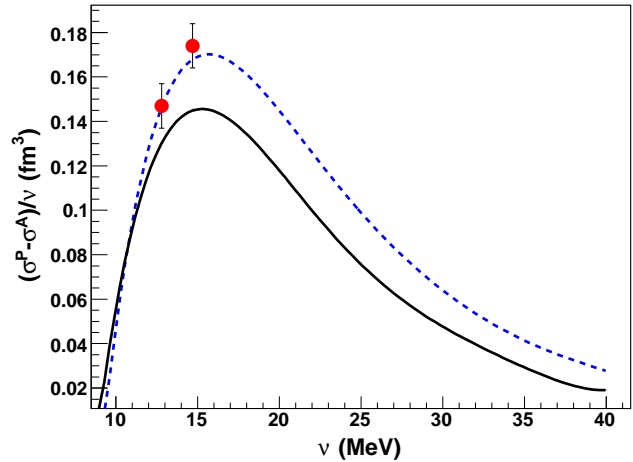}
    \caption{(Color online) GDH integrand results with statistical uncertainties only compared with the theoretical predictions from Deltuva {\it al.} (dashed curve) and Skibi\'nski {\it et al.} (solid curve).}
    \label{fig:gdh}
\end{figure}

Fig.~\ref{fig:gdh} shows the contributions from three-body photodisintegration to the $^3$He GDH integrand together with the predictions based on the computational technologies of Refs.~\cite{Deltuva2,Skibinski} as a function of the incident photon energy. Our data are in very good agreement with predictions of Deltuva {\it et al.} Both predictions show that the GDH integrand maximizes at 16 MeV and decreases significantly after 40 MeV. As such, extending these measurements to higher photon energies and carrying out measurements on two-body breakup channel will provide crucial tests of the differential cross sections, the energy dependence of the predictions, and whether the contribution to the GDH integral is indeed dominated by the three-body channel below the pion threshold. These measurements, when combined with data above pion threshold from other laboratories, will directly test the $^3$He GDH sum rule prediction. They will also provide a unique test of how effective a polarized $^3$He target is a polarized neutron target. Such efforts are in progress.

The authors thank W. Chen, R. Lu and X. Zong for their help during the initial stage of this experiment, M. Souza and G. Cates for their assistance in constructing the Sol-Gel coated cell, and the TUNL personnel for the technical support of this experiment. This work is supported by the U.S. Department of Energy under contract numbers DE-FG02-03ER41231, DE-FG02-97ER41033, DE-FG02-97ER41041, Duke University, and the Polish National Science Center under Grant No. DEC-2011/01/B/ST2/00578. The numerical calculations of Krak\'ow theoretical group have been performed on the supercomputer clusters of the JSC, J\"ulich, Germany.

\end{document}